      [1999/12/01 v1.4c Il Nuovo Cimento]
\documentclass{cimento}

\usepackage{graphicx}  % got figures? uncomment this
\title{GRB environment properties through X and Optical Afterglow observations}
\author{M.L.~Conciatore\from{ins:1}\ETC,
L.A.~Antonelli\from{ins:2},
G.~Stratta\from{ins:3},
F.~Fiore\from{ins:2}\\
        \atque
R. Perna \from{ins:4} }
\instlist{\inst{ins:1} Universit\'a degli Studi di Roma "La Sapienza",
INAF-Osservatorio Astronomico di Roma \inst{ins:2} INAF-OAR
\inst{ins:3}INAF-OAR $\&$ OMP \inst {ins:4} JILA-Boulder}
\PACSes{\PACSit{98.70.Rz}{gamma-ray sources; gamma-ray bursts}
\PACSit{98.58.Ca}{Interstellar dust grains}}
\begin{document}

\maketitle

\begin{abstract}
We present the spectral analysis of 14 gamma-ray bursts (GRB) X-ray
   afterglows in order to investigate the properties of interstellar
   matter (ISM) along the line of sight of GRB. We carried out a
   simultaneous analysis of the NIR-optical and X-band for those
   afterglows with an optical counterpart too, in order to evaluate
   and strongly constrain the absorption effect on the spectral energy
   distribution due to dust extinction from GRB environment. We
   evaluated the equivalent hydrogen column density $N_H$ from X-ray
   spectroscopy and rest frame visual extinction $A_V$ by assuming
   different type of ISM composition and dust grain size
   distribution. From our analysis we obtained a distribution of the
   GRB rest frame consistent with the one expected if
   GRB were embedded in a Galactic-like molecular cloud. Moreover,
   values of the visual extinction estimated from the simultaneous
   analysis of NIR-to-X band favour an environment where small dust
   grain are destroyed by the interaction with the X-ray and UV
   photons from GRB.
%This sample paper is intended to briefly expose the differences
%between a standard \LaTeX\ article and a paper based upon the
%\texttt{cimento} class.
\end{abstract}

\section{Introduction}
It's now generally believed that long-duration GRB are associated with the
collapse of massive stars (e.g. \cite{ref:woo}) and so a dense and dusty
environment, typical of star forming regions, is expected in the nearby of the
bursts. Up to now there are some observations that are in good agreement with
this molecular cloud-like scenario, like the emitting and absorbing
features observed in some X-ray afterglows (e.g. \cite{ref:anto},
\cite{ref:piro}), the detection of a large amount of
dust obtained from the high resolution spectroscopy of three optical
 afterglows \cite{ref:sava} and the high rest frame visual extinction
 inferred in some other optical afterglows (e.g.\cite{ref:bloom98}).
 Under this point of view, the non detection of $\sim40\%$ of X-ray
 afterglows at optical wavelengths (dark GRB) would be due to the dust
 absorption. On the other hand, the spectral energy distribution of
 some GRB afterglows indicates that dust reddening is very slow (
 \cite{ref:simon01},\cite{ref:GalamaWijers}). Moreover, for the
 observed optical afterglows, on average, the estimated rest frame
 visual extinction is a factor 10-100 lower than the expected if an
 ISM with dust to gas ratio and an extinction curve similar to the
 Galactic one are assumed (\cite{ref:prede}). Also, some high redshift
 burst showed a larger amount of gas than dust. In a recent work,
 Stratta et al.(\cite{ref:giulia}) already pointed out that absorption
 properties derived from X-ray and optical afterglows spectral
 analysis may require a ``non-standard'' extinction. A more careful
 study is required to probe the ISM of GRB environment.\\ In this work
 we present the results of a systematic multiwavelenght spectra
 analysis of a sample of GRB afterglows in order to put some
 constraints on the properties of the circumburst environment.\section{X-ray Data Analysis}

\begin{table}
\begin{minipage}[ht]{\columnwidth}
\centering
\caption{\small X-ray afterglows sample. $^{a}$Galactic
Coordinates (J2000), $^{b}$Redshift, obtained through optical
  spectroscopy, $^{c}$Optical Transient, $^{d}$Galactic hydrogen
  column density along the line of sight of each GRB (from Dickey
  $\&$ Lockman maps \cite{ref:DLmap}) }
\label{campione}
\begin{tabular}{|cccccc|}  
{\bf GRB} & {\bf $^{a}R.A.$}& {\bf $^{a} Dec$} & {\bf$^{b}Z$}&{\bf$^{c}OT$}& {\bf $^{d}N^{gal}_{H}$ }\\
{\bf } & { \small (h m s)} & { \small ( $^{\circ}$ ' '')} &{}&{} &{\small ($10^{22} cm^{-2}$)}\\ 
\hline
{\small \bf 991216 } &05 09 31 &+11 16 50  &1.02 &yes &0.27 \\
{\small \bf 000210 } &01 59 13&-40 40 46&0.846&no&0.0223\\
{\small \bf 000926 } &17 04 10 &+51 46 32  &2.0375 &yes &0.0265\\
{\small \bf 001025 } &08 36 36.5 &-13 04 30.3  &1&no&0.065 \\
{\small \bf 011211 } &11 15 16.4 &-21 55 44.8  &2.14&yes&0.0427 \\
{\small \bf 020322 } &18 00 53.0&+81 04 48.0   &1& no&0.045 \\
{\small \bf 020405 } &13 57 54 &-31 23 34  &0.691 &yes& 0.0427  \\
{\small \bf 020813 } &19 46 41  & -19 36 00 &1.254 &yes&0.075  \\
{\small \bf 021004 } &00 26 54 &+18 55 50 &2.328 &yes&0.0427 \\
{\small \bf 030226 } &11 33 03 &+25 54 20  &1.986&yes &0.0181  \\
{\small \bf 030227 } &04 57 29.0 &20 29 23.9  &1.6& yes&0.218  \\
{\small \bf 031203 } &08 02 30.0 &-39 50 48.0 &1 &no&0.621  \\
{\small \bf 040106 } &11 58 50.5 &-46 47 14.0 &1&no&0.0842  \\
{\small \bf 040223 } &16 39 34.0&-41 55 45.0 &1 &no&0.663  \\
\end{tabular} 
\end{minipage}  
\end{table}

 We selected from the XMM-Newton and Chandra archives all the X-ray
afterglows with an high signal-to-noise ratio (see Tab \ref{campione})
in order to perform a good spectral analysis. Standard data reduction
in the 0.1-10.0 keV energy range was performed using SAS 6.0 for the
data of the XMM-Newton EPIC instrument and using CIAO 2.3 for the
Chandra ACIS-S instrument. All the spectra were analyzed with Xspec
11.2.0. In order to get Gaussian statistics, thus so to ensure the
applicability of the $\chi^2$ test to evaluate the goodness of our
fits, the spectra were rebinned to obtain at least 20 counts per
energy channel. For the EPIC spectra we performed simultaneous PN,
MOS1 and MOS2 spectral fitting.
\subsection{The spectral model}
According to the standard fireball model, the spectrum of the X-ray
afterglows emission is well described by a simple power law
$f_{X}(E)\propto E^{-\beta_{X}}$. In this analysis we adopted a
spectral model that takes into account also for the photoelectric
absorption both Galactic and extragalactic due to the metal rich
material along the line of sight of GRB:
$f_{X}\propto\left(E^{-\beta_{X}}\times e^{-N^{gal}_{H}}\times
e^{-N^{Z}_{H}}\right)$, where $N^{gal}_{H}$ is the is the equivalent
hydrogen Galactic column density fixed according to the Dickey $\&$
Lockman map \cite{ref:DLmap} and $N^{Z}_{H}$ is the extragalactic
contribution to the absorption, whose value was left free to vary. The
redshift values are derived from optical observation when available;
otherwise, the value z=1 was adopted, being this the peak value of GRB
redshift distribution\cite{ref:DJ01}.
\subsection{Results}
\begin{figure}[!ht]\centering\includegraphics[width=7cm]{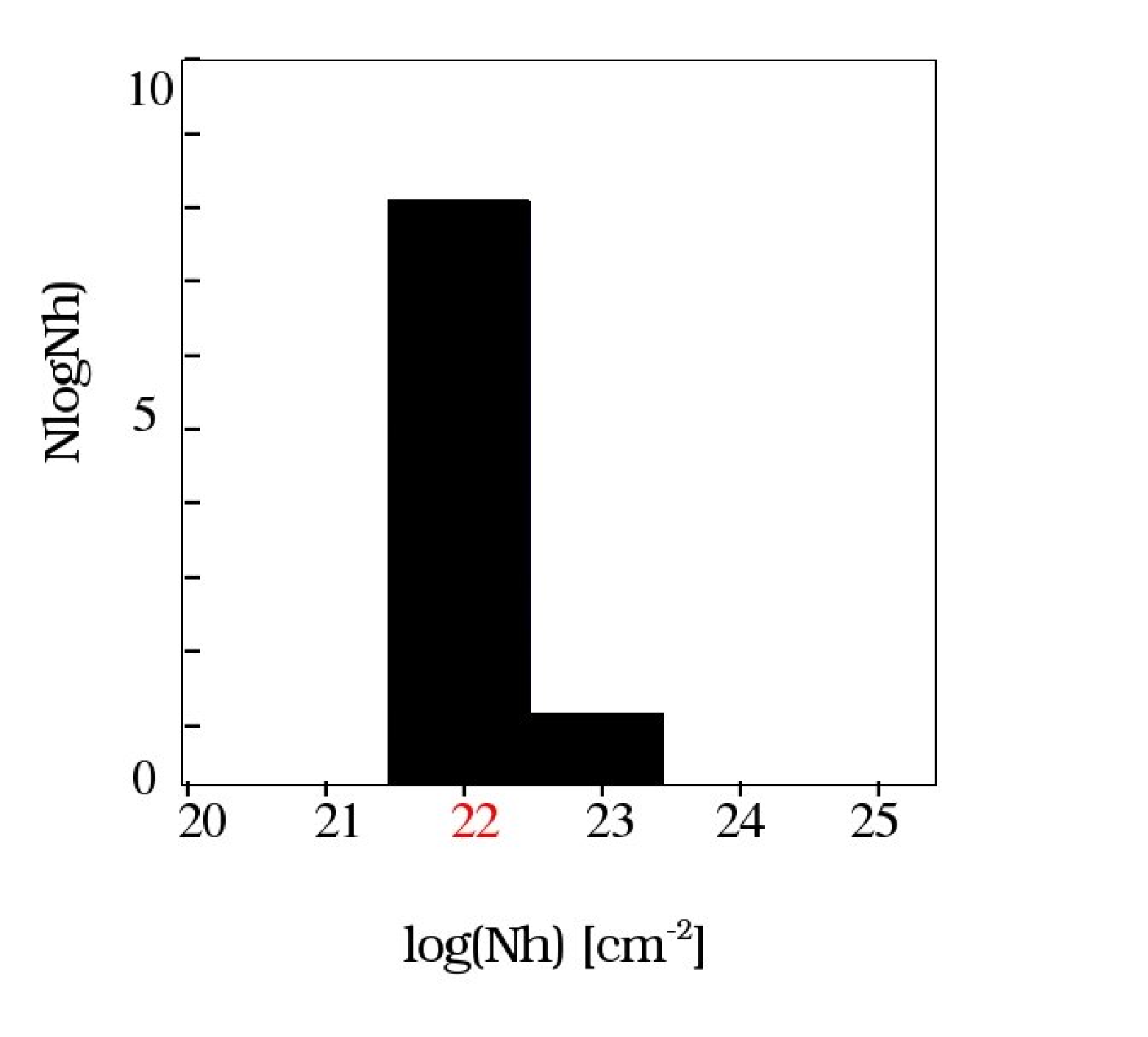}\includegraphics[width=7cm]{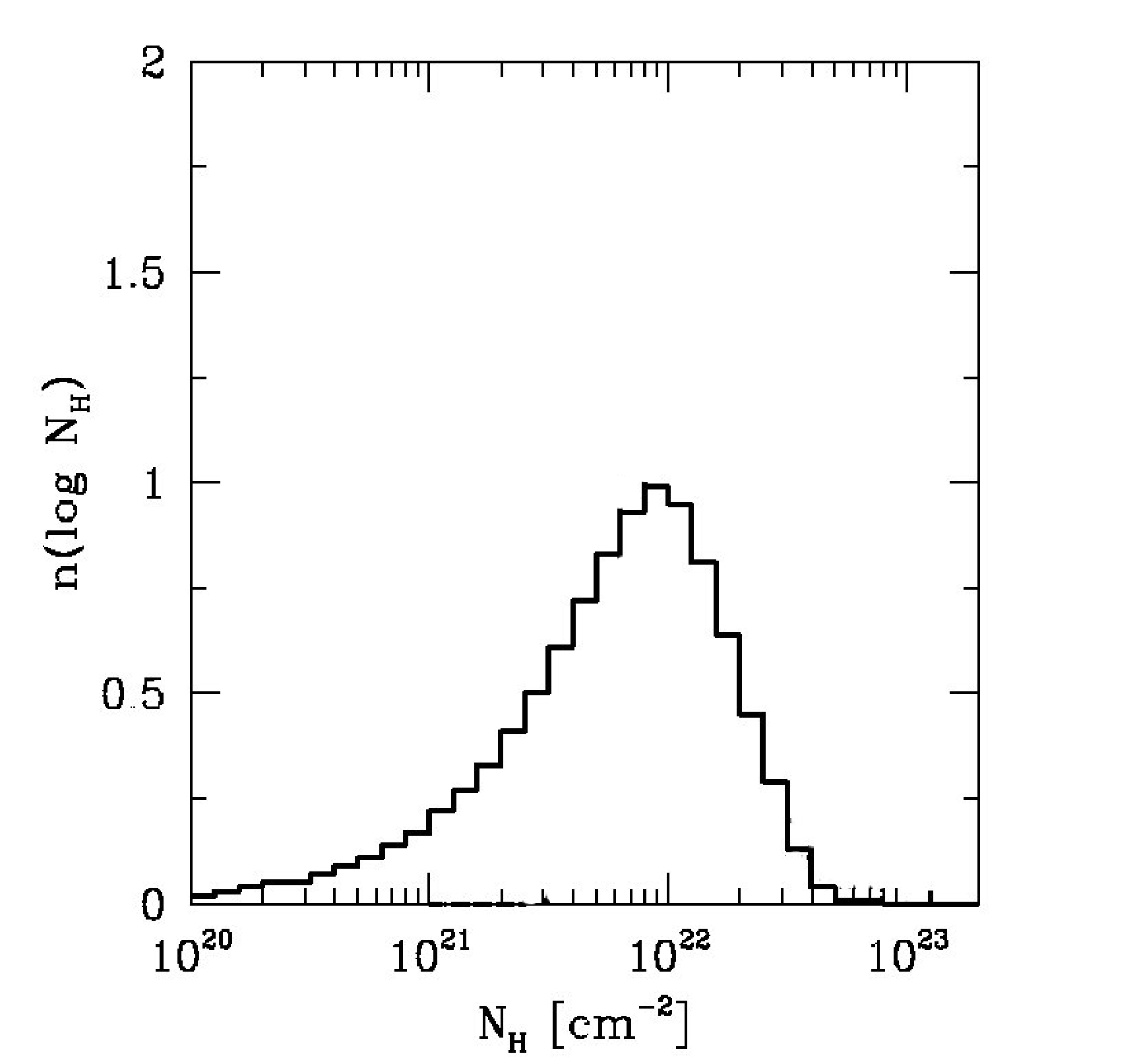}\caption{\label{results}(\it{Left}) The rest frame equivalent $N_{H}$ distribution obtained from the analysis in this work and the
 expected $N^{Gal}_{H}$ distribution for GRB occurring in Galactic-like
 molecular cloud from Reichart $\&$ Price \cite{ref:RP02} (\it {Right}). }
\end{figure}

We found an absorption larger than the Galactic value in nine cases
(GRB000210,GRB001025, GRB020322, GRB020405, GRB020813, GRB030226,\\
GRB030227, GRB031203, GRB040223) with confidence level up to
4$\sigma$. The rest frame equivalent
hydrogen column density distribution obtained has a
weighted average of $<N^{Z}_{H}>=(1.0\pm0.1)\times10^{22}cm^{-2}$,
consistent with the theoretical peak expected if GRB are embedded in
a Galactic-like molecular cloud (see Fig.\ref{results}).
\section{Multiband analysis}
For the eight afterglows of the sample with an optical counterpart
(see Tab\ref{campione}) the analysis was extended to the optical and
NIR band, in order to better constrain extinction properties
through a fit of spectra as wide as possible. Photometric data have
been taken from literature. To perform this large band analysis we
applied some correction to data. Firstly, being $F(t)\propto
t^{-\alpha}$ we extrapolated magnitudes at the same time of X-ray
observation, adopting the measured optical-NIR decay index published in
literature and then we corrected them from Galactic extinction along
GRB line of sight using the IRAS 100 m$\mu$ E(B-V) maps by Schlegel et
al. 1998
\cite{ref:SCHmap} and by deriving the extinction at different
wavelengths using the extinction curve parameterization taken from Cardelli et
al. 1989 \cite{ref:cardelli} assuming $R_{V}=A_{V}/E(B-V)=3.1$. Finally, the magnitudes have been converted in fluxes using the
effective wavelenghts and normalization fluxes given by Fukugita et
al. 1995 \cite{ref:FUKU}.
\subsection{The spectral model}
For the optical/NIR data we adopted a model composed by a powerlaw and
an absorption component that takes into account the dust extinction:
$f_{o}(\lambda)=C\lambda^{-2-\beta}e^{\left(-A(\lambda r
)/A_{Vr}\right)A_{Vr} }$, where $A_{Vr}$ is the visual
extinction. Assuming the slow cooling regime as the most probable one
for the electron population at the observation time, we assumed
that the electron index p is twice the X-ray spectral index previously
obtained within $90\%$ confidence level. We fitted the data using both
the possibility predicted for this regime: $\beta=(p-1)/2$ if $
\nu_{opt/NIR}<\nu_{c}$, $\beta=p/2$ if $
\nu_{opt/NIR}>\nu_{c}$, where $\nu_{c}$ is the cooling frequency. We tested
different dust composition and dust-to-gas ratio using different
extinction curves, assuming that the dust grains distribution $n(a)$
is described by a simple power law $dn(a)\propto a^{-q}da $. The
curves are: the Galactic-like (G) from Cardelli et al. 1989
\cite{ref:cardelli}, for which $q=-3.5$, $a_{min}=0.005\mu
m<a<a_{max}=0.25\mu m$ ; the Small Magellanic Cloud-like (SMC) from
Pei 1992 (\cite{ref:pei}), that's the same model but with 1/8 of the
solar metallicity; two extinction curves (Q1 and Q2) obtained by
Maiolino et al 2000 (\cite{ref:maiolino}), from simulation from
study of a sample of AGN, for which $q=-3.5$, $a_{min}=0.005\mu
m<a<a_{max}=10\mu m$ and $q=-2.5$, $a_{min}=0.005\mu m<a<a_{max}=1\mu
m$ respective. We have also tested an extinction curve (C) derived
by Calzetti et al. 2001 \cite{ref:Cal} for a sample of local starburst
galaxies. For all this different ISM model the rest frame visual
extinction has been estimated for all the afterglows sample.
\subsection{Results}
\begin{figure}[!h]\centering\includegraphics[width=10cm]{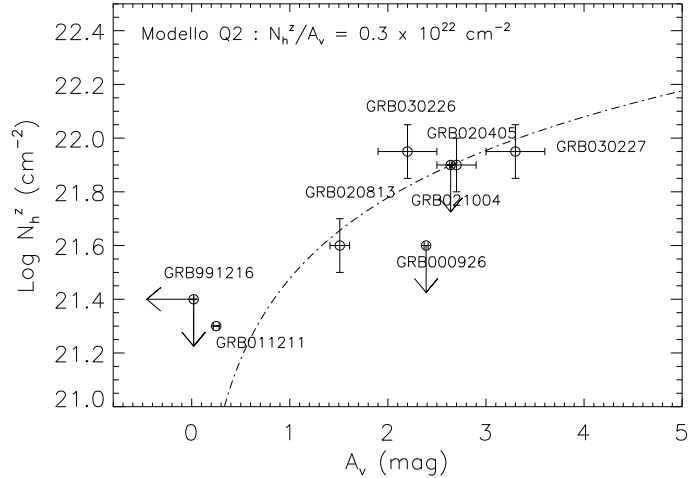}\caption{\label{results2} The best fit additional column density against best
 fit visual extinction obtained using Q2 curves.}
\end{figure}

For GRB991216 and GRB011211 we have obtained only the Galactic
contribution to extinction, a result consistent with the previously
X-ray analysis.\\ For GRB000926, GRB020405, GRB020813, GRB030226 and
GRB030227 the best fit is obtained assuming an ISM with dust grain
distribution skewed toward large grains (Q1 curve for the first case,
Q2 curve for the other four cases). We also compared the best fit
additional $N_{H}$ density at GRB's redshift with the best fit $A_{V}$
obtained for the different extinction curves. The $N_{H}/A_{V}$
relations have been compared with the corresponding theoretical one:
Galactic, $N_{H}/A_{V}=0.18\times 10^{22}~cm^{-2}$ \cite{ref:prede};
Small Magellanic Cloud, $N_{H}/A_{V}=1.6\times 10^{22}~cm^{-2}$
\cite{ref:WEINGART} ;  Q1, $N_{H}/A_{V}=0.7\times 10^{22}~cm^{-2}$
and Q2, $N_{H}/A_{V}=0.3\times 10^{22}~cm^{-2}$
\cite{ref:maiolino}. For the starburst galaxies no $N_{H}/A_{V}$
relationship has been derived due the complexity of this kind of
ISM. Assuming a Galactic-like and SMC ISM, this ratio are well above
the expected values, confirming previous studies
(e.g. \cite{ref:GalamaWijers},\cite{ref:giulia}). A good agreement is
obtained with an ISM with dust grain size distribution skewed toward
large grain (see fig. \ref{results2}). Such a dust composition
reconciles the typical low reddening observed in the optical afterglows
SEDs with the high amount of dust observed through optical
spectroscopy \cite{ref:sava}. This kind of environment can be obtained
in two ways: small dust grain can coagulate into larger producing a
dust distribution biased towards larger grains as expected in high
density medium (e.g.\cite{ref:kim}), or the physical state of the gas
and dust are modified by the intense X-ray and UV emission from GRB,
(e.g \cite{ref:waxdra}).\\ Such an uncertainty will be solved by
monitoring the very early afterglows and the possible evolution of the
extinction effects. The SWIFT satellite and robotic telescopes, such
as REM, will be very useful to provide us these informations.
\acknowledgments
For a more detailed analysis see Conciatore et al 2005 in prep.

\end{document}